\documentclass[a4paper,12pt]{article}
\usepackage{amssymb,amsmath,amsthm}
\usepackage{bm}%
\usepackage{ascmac}
\usepackage{mathrsfs}
\usepackage{stmaryrd}
\usepackage{comment}
\usepackage{xcolor}
	\colorlet{mycol}{black}
\usepackage{hyperref}
\hypersetup{
	colorlinks=true,
	citecolor=blue,
	linkcolor=blue,
	urlcolor=blue,
}

\def\nn{\notag}
\def\Z2{\mathbb{Z}_2^2}
\def\g{\mathfrak{g}}

\def\DP#1#2{\hat{#1}\cdot \hat{#2}}
\def\PH#1#2{(-1)^{\hat{#1}\cdot \hat{#2}}}
\def\osp{\mathfrak{osp}}
\def\sl{\mathfrak{sl}}
\def\TL{\mathcal{L}}

\title{
	An integrable hierarchy associated with loop extension of $\mathbb{Z}_2^2$-graded $\mathfrak{osp}(1|2)$}
\author{Naruhiko Aizawa\thanks{E-mail: {\it aizawa@omu.ac.jp} (corresponding author)}, \and 
	Ichi Fujii\thanks{E-mail: \textit{sq25482d@st.omu.ac.jp}}, \and
	Ren Ito\thanks{E-mail: \textit{sd22709y@st.omu.ac.jp}} 
}
\date{\today}

\begin{document}
	
	\maketitle
	\thispagestyle{empty}
	\begin{center}
		{\small{	\textit{Department of Physics, Graduate School of Science,
					\\
					Osaka Metropolitan University, Sugimoto Campus,
					\\
					Osaka 558-8585, Japan.}
		}}
	\end{center}

\vfill
\begin{abstract}
	A hierarchy of $\mathbb{Z}_2^2$-graded integrable equations is constructed using the loop extension of the $\mathbb{Z}_2^2$-graded Lie superalgebra $\mathfrak{osp}(1|2)$.
	This hierarchy includes $\mathbb{Z}_2^2$-graded extensions of the Liouville, sinh-Gordon, cosh-Gordon, and, in particular, the mKdV equations.
	The $\mathbb{Z}_2^2$-graded KdV equation is also derived from its mKdV counterpart via the Miura transformation.
	We present explicit formulas for the conserved charges of the $\mathbb{Z}_2^2$-KdV and $\mathbb{Z}_2^2$-mKdV equations.
	A distinctive feature of these $\mathbb{Z}_2^2$-graded integrable systems is the existence of conserved charges with nontrivial grading.
\end{abstract} 
	
\clearpage
\setcounter{page}{1}
%%%%%%%%%%%%%%%%%%%%%%%%%%%%%%%%%%%%%%%%%%%%%%%%%%%%%%%%%%%%%%%%%%%%%%%%%
\section{Introduction} 

This paper is a continuation of our project aiming to construct $\mathbb{Z}_2^2$-graded integrable systems. Here,``$\mathbb{Z}_2^2$-graded" (i.e., $\mathbb{Z}_2 \times \mathbb{Z}_2$-graded) means that all field variables are $\mathbb{Z}_2^2$-graded commutative (see \S \ref{SEC:algebra} for the precise definition).  
Physically, this implies that the fields describe paraparticles$-$particles beyond bosons and fermions. 
One motivation for this project comes from the recent revival of interest in paraparticles.   Several studies have investigated in detail {\color{mycol}their physical distinguishability from ordinary bosons and fermions} \cite{toppan2021inequivalent,toppanMulti,toppan2022first,toppan2024detectability,toppan2024braid,wang2025particle}. There have also been experimental attempts to realize {\color{mycol}them} in laboratory settings using trapped ions \cite{huerta2025particle,alderete2025experimental}.

The first example of a $\Z2$-graded integrable system was the $\Z2$-graded extension of the sine-Gordon equation constructed by Bruce \cite{BruSG}. This equation possesses infinitely many conserved quantities, which are obtained using the auto-B\"acklund transformation. 

Subsequently, $\Z2$-graded extensions of the Liouville and sinh-Gordon equations were obtained by generalizing the Toda field theory to the $\Z2$-graded setting \cite{AIKTTslint,AFITTsuper}. 
Toda field theory is a powerful framework for constructing integrable systems based on Lie (super)algebras. It also provides a method for finding analytic solutions using representation theory (see for example \cite{LeSa1,LeSa2,LeSa3,babelon1990conformal,toppan1991superliouville,toppanZhan1992}). 
In the $\Z2$-graded setting, the Lax operators take values in a $\Z2$-graded Lie (super)algebra. Depending on the choice of the underlying $\Z2$-graded Lie (super)algebra, one obtains inequivalent $\Z2$-graded extensions of the classical Toda equations. 
In \cite{AIKTTslint}, the Lax operators are constructed from the $\Z2$-graded extension of $\mathfrak{sl}(2)$ and its affinization.  
The resulting integrable differential equations are the $\Z2$-Liouville and $\Z2$-sinh-Gordon equations, respectively. 
In contrast, the $\Z2$-graded extension of $\osp(1|2)$ (denoted by $\Z2$-$\osp(1|2)$) is used as the basic building block in \cite{AFITTsuper}, leading to a $\Z2$-graded version of the super-Liouville equation. 
{\color{mycol}
Furthermore, in \cite{AIKTTslint,AFITTsuper}, 
the current algebras associated with the $\Z2$-(super)Liouville equations were also investigated using Polyakov’s soldering procedure \cite{polyakov1990gauge}.}
Reflecting the conformal nature of the $\Z2$-(super)Liouville equations, $\Z2$-graded extensions of the (super)Virasoro algebra were obtained as $\Z2$-graded Poisson–Lie algebras. 

One of the most important classes of equations in the study of integrable systems is the family of KdV-type equations. However, their $\Z2$-graded extensions have not yet been studied. The purpose of the present work is to construct a hierarchy of $\Z2$-graded integrable equations that includes KdV-type equations. To this end, we employ the loop extension of $\Z2$-$\osp(1|2)$ together with the Lax operator formalism. 
Our choice of Lax operators leads to a hierarchy that includes $\Z2$-graded extensions of the Liouville, sinh-Gordon, cosh-Gordon, and mKdV equations. The $\Z2$-Liouville equation obtained in this work coincides with that in \cite{AFITTsuper}. 
Our $\Z2$-mKdV equation exhibits an interesting feature: the Miura transformation can be read off directly from the equation. Using this Miura transformation, we also derive the corresponding $\Z2$-KdV equation.

We also discuss the conserved quantities associated with the integrable hierarchy. Owing to the $\Z2$-graded nature of the system, the standard procedure for deriving conserved charges cannot be applied directly. We therefore present a modified version of this procedure and provide illustrative computations for the $\Z2$-KdV and $\Z2$-mKdV equations. 
We observe that there exist conserved charges with nontrivial grading.

The paper is organized as follows. In the next section, we recall the definition of $\Z2$-graded Lie superalgebras and introduce the loop extension of  $\Z2$-$\osp(1|2).$ 
In \S \ref{SEC:formalism}, we present the methods for constructing integrable hierarchies and for computing the associated conserved quantities. 
In sharp contrast to previous works, our construction does not require superfields.
We apply this framework to the loop $\Z2$-$\osp(1|2)$ in \S \ref{SEC:hier}. 
The $\Z2$-graded Liouville, sinh-Gordon and cosh-Gordon equations are obtained from the negative hierarchy, while the $\Z2$-mKdV equation arises from the positive hierarchy. 
\S \ref{SEC:MiuraCon} consists of two subsections. 
First, we derive the $\Z2$-KdV equation from the $\Z2$-mKdV equation via the Miura transformation. The Lax operators for the $\Z2$-KdV equation is also obtained by a gauge transformation associated with the Miura transformation. 
In the second subsection,  we compute the conserved charges for $\Z2$-KdV and $\Z2$-mKdV equations using the method of \S\ref{SEC:formalism}. 
Finally, in \S \ref{SEC:Conclusion}, we summarize the results and offer some remarks on desirable and possible future developments.

%%%%%%%%%%%%%%%%%%%%%%%%%%%%%%%%%%%%%%%%%%%%%%%%%%%%%%%%%%%%%%%%%%%%%%%%%
\section{Loop extension of $\Z2$-graded $\osp(1|2)$} \label{SEC:algebra}
\setcounter{equation}{0}

 Let us recall the definition of the $\Z2$-graded Lie superalgebra \cite{rw1,rw2,scheu}. 
Let $\mathfrak{g}$ be a vector space and $\hat{a}\equiv [a_1a_2]$ an element of $\Z2$. 
Suppose that $\mathfrak{g}$ is a direct sum of graded components 
\begin{equation}
	\mathfrak{g} = \bigoplus_{\hat{a} \in \Z2}  \g_{\hat{a}}= \g_{[00]} \oplus \g_{[10]} \oplus \g_{[01]} \oplus \g_{[11]}.   
\end{equation}
If $\mathfrak{g}$ admits a bilinear operation (the graded Lie bracket), denoted by $ \llbracket \cdot, \cdot \rrbracket $ and satisfying the identities 
\begin{align}
	& \llbracket A_{\hat{a}}, B_{\hat{b}}  \rrbracket \in \g_{\hat{a}+\hat{b}},
	\\
	& \llbracket A_{\hat{a}}, B_{\hat{b}} \rrbracket = -(-1)^{\hat{a}\cdot\hat{b}} \llbracket  B_{\hat{b}}, A_{\hat{a}} \rrbracket,
	\\
	& (-1)^{\hat{a}\cdot\hat{c}} \llbracket A_{\hat{a}}, \llbracket B_{\hat{b}}, C_{\hat{c}} \rrbracket \rrbracket + \PH{b}{a} \llbracket B_{\hat{b}}, \llbracket C_{\hat{c}}, A_{\hat{a}} \rrbracket \rrbracket + \PH{c}{b}\llbracket C_{\hat{c}}, \llbracket A_{\hat{a}}, B_{\hat{b}} \rrbracket \rrbracket  = 0, \label{Jacobi}
\end{align}
where $ A_{\hat{a}}, B_{\hat{a}}, C_{\hat{a}} $ are homogeneous elements of $\mathfrak{g}_{\hat{a}}$ and 
\begin{equation}
	\hat{a} + \hat{b} = [a_1+b_1 ,a_2+b_2] \in \Z2, \qquad 
	\hat{a} \cdot \hat{b} = a_1 b_1 + a_2 b_2 \in \mathbb{Z}_2,
\end{equation}
then $\mathfrak{g}$ is referred to as a $\Z2$-graded Lie superalgebra.

It is clear from the definition that the graded Lie brackets are realized by commutators and anticommutators as follows 
\begin{equation}
	\llbracket A_{\hat{a}}, B_{\hat{b}} \rrbracket
	= 
	\begin{cases}
		[A_{\hat{a}}, B_{\hat{b}}], & \DP{a}{b} = 0,
		\\[10pt]
		\{ A_{\hat{a}}, B_{\hat{b}} \}, & \DP{a}{b} = 1.
	\end{cases}
\end{equation}
If $\llbracket A, B \rrbracket = 0, $ we say that $A$ and $ B$ are $\Z2$-\textit{graded commutative}. 
It is also observed from the definition that $ \g$ has $\mathbb{Z}_2$-grading, too:
\begin{equation}
	\g = \g_{[0]} \oplus \g_{[1]}, \qquad \g_{[0]} := \g_{[00]} \oplus \g_{[11]}, \ 
	\g_{[1]} := \g_{[10]} \oplus \g_{[01]}.
\end{equation}

In the present work, we take the $\Z2$-graded extension of the Lie superalgebra $\osp(1|2)$ introduced in \cite{Aizawa_2020,AmaAi} and consider its loop extension. 
This infinite-dimensional $\Z2$-graded loop superalgebra is also denoted by $\g$; this should not cause any confusion. Each $\Z2$-graded subspace of $\g$ is also infinite-dimensional, and they are given by
\begin{equation}
	\g_{[00]} = \{ \ K^0_m, \ K^{\pm}_m \ \}, 
	\quad
	\g_{[10]} = \{  P^{\pm}_m \ \}, 
	\quad  
	\g_{[01]} = \{  Q^{\pm}_m \ \}, 
	\quad
	\g_{[11]} = \{ \ L^0_m, \ L^{\pm}_m \ \}
\end{equation}
where $m \in \mathbb{Z}.$ 
We now present their non-vanishing relations in terms of (anti)commutators. 
The $\g_{[0]}$ sector forms a Lie subalgebra	
\begin{alignat}{3}
	[K^0_m, K^{\pm}_n] &= \pm 2 K^{\pm}_{m+n}, & \qquad 
	[K^+_m, K^-_n] &= K^0_{m+n}, & \qquad 
	[L^0_m, L^{\pm}_n] &= \pm 2 K^{\pm}_{m+n},
	\nn \\
	[L^+_m, L^-_n] &= K^0_{m+n}, & 
	[K^0_m, L^{\pm}_n] &= \pm 2 L^{\pm}_{m+n}, &
	[K^{\pm}_m, L^{\mp}_n] &= \pm L^0_{m+n},
	\nn \\
	[L^0_m, K^{\pm}_n] &= \pm 2 L^{\pm}_{m+n}.
\end{alignat}
The relations between $ \g_{[0]}$ and $ \g_{[1]}$ are determined by commutators and anticommutators:
\begin{alignat}{3}
	[K^0_m, P^{\pm}_n] &= \pm P^{\pm}_{m+n}, &\qquad    
	[K^{\pm}_m, P^{\mp}_n] &= -P^{\pm}_{m+n}, &\qquad
	[K^0_m, Q^{\pm}_n] &= \pm Q^{\pm}_{m+n},
	\nn \\[3pt]
	[K^{\pm}_m, Q^{\mp}_n] &= -Q^{\pm}_{m+n}, & 
	\{ L^0_n, P^{\pm}_m  \} &= \pm i Q^{\pm}_{m+n}, & 
	\{  L^{\mp}_n, P^{\pm}_m \} &= -i Q^{\mp}_{m+n},
	\nn \\[3pt]
	\{ L^0_n, Q^{\pm}_m \} &= \mp i P^{\pm}_{m+n},  & 
	\{ L^{\mp}_n, Q^{\pm}_m \} &= i P^{\mp}_{m+n}
\end{alignat}
and the same applies to the $\g_{[1]}$-sector 
\begin{alignat}{3}
	\{ P^{\pm}_m, P^{\pm}_n \} &= \pm 2 K^{\pm}_{m+n}, &\qquad 
	\{P^+_m, P^-_n \} &= K^0_{m+n}, & \qquad 
	[P^{\pm}_m, Q^{\pm}_n] &= \pm 2i L^{\pm}_{m+n},
	\nn \\[3pt]
	[P^{\pm}_m, Q^{\mp}_n] &= i L^0_{m+n}, &
	\{ Q^{\pm}_m, Q^{\pm}_n\} &= \pm 2 K^{\pm}_{m+n}, & \{Q^+_m, Q^-_n\} &=K^0_{m+n}.
\end{alignat}
The elements with $m = 0 $ correspond to the ten-dimensional $\Z2$-graded Lie superalgebra \cite{Aizawa_2020,AmaAi}
\begin{equation}
 \Z2\text{-}\osp(1|2) = \{ \ K^0_0, \ K^{\pm}_0,\ P^{\pm}_0, \ Q^{\pm}_0, \ L^0_0, \ L^{\pm}_0 \  \}. 
 \label{FiniteDalg}
\end{equation}

One may introduce $\Z2$-graded derivation $ d_{\hat{m}}$ for $ \hat{m} \in \Z2$ by  
\begin{equation}
	d_{\hat{m}}(\llbracket A_k, B_n \rrbracket) = 
	\llbracket d_{\hat{m}}(A_k), B_n \rrbracket + \PH{m}{a} \llbracket A_k, d_{\hat{m}}(B_n) \rrbracket, 
	\quad A_n \in \g_a, \ B_k \in \g_b
\end{equation}
It was shown in \cite{AiSe} that $\g$ admits the [00] and [11]-graded derivations, but none of [10], [01]-graded. 
The [00]-graded derivation counts the mode as usual {\color{mycol}(Recall that $ X_m := \lambda^m \otimes X$ where $ X \in \Z2\text{-}\osp(1|2)$)}:
\begin{equation}
	[d_{00}, X_m] = mX_m, \quad X_m \in \g
\end{equation}
The action of [11]-graded derivation is determined by
\begin{alignat}{3}
	[d_{11}, K_m^0] &= m L^0_m, & \qquad [d_{11}, L_m^0] &= m K_m^0, & \qquad [d_{11}, K_m^{\pm}] &= m L^{\pm}_m,
	\notag \\[3pt]
	[d_{11}, L_m^{\pm}] &= m K_m^{\pm}, & \{d_{11}, P_m^{\pm}\} &= i m Q_m^{\pm},  & \{ d_{11}, Q_m^{\pm}\} &= -im P_m^{\pm},
	\notag \\[3pt]
   [d_{11}, d_{00}] & = 0.
\end{alignat}
%As usual, $d_{00}$ count the power of the spectral parameter $\lambda$:
%$ X_m := \lambda^m \otimes X$ where $ X \in \Z2\text{-}\osp(1|2)$.

The $\Z2$-loop superalgebra $\g$ admits various gradations. 
The derivation $d_{00}$ defines the homogeneous gradation:
\begin{align}
	\g = \bigoplus_{n \in \mathbb{Z}} \g_n, \qquad 
	\g_m = \{\  X \in \g \ \left| \ [d_{00}, X] = m X \right. \ \}. \label{HomoG}
\end{align}
Combining this with the intrinsic $\Z2$-gradation, one obtains a bi-gradation $ \Z2 \times \mathbb{Z}.$ As an alternative single gradation, we consider the operator defining the principal gradation of $\widehat{\sl}(2)$:
\begin{equation}
	\mathcal{G} = \frac{1}{2} K^0_0 + 2 d_{00}.
\end{equation}
This [00]-graded operator induces a  half-integer gradation on $\g$:
\begin{align}
	\g = \bigoplus_{m \in \mathbb{Z}} \g_{m/2}, \qquad 
	\g_p = \{ \ X \in g \ \left| \ [\mathcal{G},X] = p X \right. \ \}. \label{PrincipleG}
\end{align}
More precisely, $\g$ decomposes into four types of subspaces:
\begin{align}
	\g_{2m} &=\mathrm{Span}\langle \ K_m^0,\; L_m^0 \ \rangle, \nn \\
	\g_{2m\pm\frac{1}{2}} &=\mathrm{Span}\langle \ P_m^\pm,\; Q_m^\pm \ \rangle, \nn \\
	\g_{2m+1} &=\mathrm{Span}\langle \ K_m^+,\; K_{m+1}^-,\; L_m^+,\; L_{m+1}^- \ \rangle.
	\label{HIgradesSp}	
\end{align}
The positive part $ \bigoplus_{ p > 0} \g_p $ is generated by the elements in $ \g_{1/2} \oplus \g_1$:
\[
  \g_{1/2} = \{  \  P^+_0, \; Q^+_0 \ \}, \qquad 
  \g_1 = \{ \  K^+_0,\; K^-_1, \; L^+_0, \; L^-_1  \  \}.
\]
Similarly, the negative part $ \bigoplus_{ p < 0} \g_p $ is generated by the elements in $ \g_{-1/2} \oplus \g_{-1}$:
\[
  \g_{-1/2} = \{  \  P^-_0, \; Q^-_0 \ \}, \qquad 
  \g_{-1} = \{ \  K^+_{-1},\; K^-_0, \; L^+_{-1}, \; L^-_0  \  \}.
\]
We will use this gradation to construct an integrable hierarchy. 

Finally, we remark that $\g$ admits two central extensions, one of which is [00]-graded and the other [11]-graded \cite{AiSe}.  Here the central elements $\Z2$-commute with all elements of $\g$ as well as with  the derivations. 
These central extensions endow $\g$ with the structure of a  $\Z2$-graded affine Lie superalgebra:
\[
  \Z2\text{-}\widehat{\osp}(1|2) = \g \oplus \mathbb{C} d_{00} \oplus \mathbb{C} d_{11} \oplus \mathbb{C} c_{00} \oplus \mathbb{C} c_{11}.
\]

%%%%%%%%%%%%%%%%%%%%%%%%%%%%%%%%%%%%%%%%%%%%%%%%%%%%%%%%%%%%%%%%%%%%%%
\section{Formalism} \label{SEC:formalism}
\setcounter{equation}{0}

\subsection{Construction of integrable hierarchy}\label{SEC:formHi}

We employ a well-developed method to construct a classical integrable hierarchy. 
There are many publications on this topic; see, for example, \cite{aratyn2004algebraic,Aratyn2001,Ferreira} and references therein. 
We denote the coordinates of the two-dimensional spacetime by $ x_{\pm},$ although they do not necessarily represent light-cone coordinates. Their interpretation depends on the context.
The integrability of the resulting equations is ensured by their zero-curvature (Lax) representation:
\begin{equation}
	[\partial_+-\mathcal{L}_+,\partial_--\mathcal{L}_-]=0, \quad \partial_{\pm} := \frac{\partial}{\partial x_{\pm}}. 
	\label{0curvEq}
\end{equation}
$\mathcal{L}_{\pm}$ contain some functions of $x_{\pm}$ and take values in the $\Z2$-graded loop  superalgebra $\g$ introduced in the previous section. We mimic the construction of the Toda hierarchy for $\widehat{\sl}(2)$. 
Using the {\color{mycol}principal} gradation of $\g$ given in \eqref{PrincipleG}, we fix $ \mathcal{L}_+ $ in the following form
\begin{align}
	\mathcal{L}_+ &=A_0+A_{\frac{1}{2}}+E_1,
	\\
	A_0 &= B^{-1} \partial_+ B \in \g_0 \quad \text{where} \quad  B:= \exp(\varphi_{00} K^0_0 + \varphi_{11} L^0_0),
	\nn \\
	A_{\frac{1}{2}} &=\sigma_{10}P_0^++\sigma_{01}Q_0^+ \in \g_{1/2}
	\nn \\
	E_1 &=K_0^++K_1^- \in \g_1 \label{LxDefinition}
\end{align}
%One may include more basis and more fields, however  
{\color{mycol}
As we will see, this  choice gives rise to hierarchy that includes $\Z2$-graded extensions of the Liouville, sinh-Gordon, cosh-Gordon, and mKdV equations. 
If we choose another $\mathcal{L}_x$, then we will obtain a different hierarchy. 
}

Here, the $ \varphi$'s and $\sigma$'s are $\Z2$-graded commutative functions of $ x_{\pm}$  and their $\Z2$-gradings are indicated by the subscripts. 
More precisely, they satisfies the following relations:
\begin{align}
	[\varphi_{00}, \varphi_{11}] &= [\varphi_{00}, \sigma_{10}] = [\varphi_{00}, \sigma_{01}] = 0,
	\nn\\
	\{ \varphi_{11}, \sigma_{10} \} &= \{ \varphi_{11}, \sigma_{01} \} = [\sigma_{10}, \sigma_{01}] = 0,
	\nn \\
	\sigma_{10}^2 &= \sigma_{01}^2 = 0.
\end{align}
These functions will serve as the dynamical fields of the {\color{mycol}integrable hierarchy}.

We consider various combinations of the half-integer graded subspaces \eqref{HIgradesSp} for $\mathcal{L}_-$, which give rise to a hierarchy of integrable systems.
For instance, we consider a combination of positive graded subspaces
\begin{equation}
	\mathcal{L}_- = D_0 + D_{1/2} + D_1 + \cdots + D_N, 
\end{equation}
where $ D_p \in \g_p$ is a liner combination of elements in $\g_p$ and $N$ is a positive half-integer. 
Since $\mathcal{L}_-$ is not graded, the coefficients of the linear combinations must be $\Z2$-graded. We suppose that the coefficients are $\Z2$-commutative and may depend on the spacetime coordinates. 
Then the zero-curvature equation \eqref{0curvEq} is decomposed into a system of equations according to the gradation:
\begin{alignat}{2}
	&\text{grade } N+1: &  
	[E_1, D_N] &= 0,
	\nn \\
	&\text{grade } N+\frac{1}{2}: & 
	[A_{1/2}, D_N] + [E_1, D_{N-1/2}] &= 0,
	\nn \\
	&\text{grade } N: & \qquad 
	-\partial_+ D_N + [A_0, D_N] + [A_{1/2}, D_{N-1/2}] + [E_1, D_{N-1}] &= 0,
	\nn \\
	& \qquad \vdots &  &\  \vdots
	\nn \\
	&\text{grade }1: & -\partial_+ D_1 + [A_0, D_1] + [A_{1/2}, D_{1/2}] + [E_1, D_0] &= 0,
	\nn \\
	&\text{grade }\frac{1}{2}: & \partial_- A_{1/2} - \partial_+ D_{1/2} + [A_0, D_{1/2}] + [A_{1/2}, D_0] &= 0,
	\nn \\
	&\text{grade } 0: & \partial_- A_0 - \partial_+ D_0 + [A_0, D_0] &= 0.
\end{alignat}
We have $2N+3$ equations for $2N+1$ yet undetermined $D$'s. 
The highest grade equation constrains the coefficients in $ D_N. $ The grade $N+\frac{1}{2}$ equation relates the coefficients in $ D_{N-1/2}$ and $ D_N$ to $ \sigma_{10}, \sigma_{01}. $ 
Proceeding in this manner, by solving the equations successively from the highest grade down to the lowest, one may derive the desired equations of motion.

Instead of using the positive graded subspaces, one may carry out the same procedure for the negative graded subspaces.
Furthermore, it is known that this procedure can also be applied to mixtures of positive and negative graded subspaces \cite{MixedHi}.

%%%%%%%%%%%%%%%%%%%%%%%%%%%
\subsection{Conserved quantities}
\label{SEC:Charges}

Given Lax operators with a spectral parameter, it is, in principle, possible to derive infinitely many conserved quantities. 
Such Lax operators is obtained from  an irreducible representation of the underlying finite dimensional $\Z2$-graded Lie superalgebra  $\Z2$-$\osp(1|2)$. 
We employ the six-dimensional representation of $\Z2$-$\osp(1|2)$ defined on a $\Z2$-graded vector space \cite{AFITTsuper}. 
The basis of the representation space is ordered according to the $\Z2$-grading as 
$ [00], \; [11], \; [01], \; [10] $ so that the representation matrices have a block structure. 
Each block has a fixed $\Z2$-grading as follows
\[
\begin{bmatrix}
	A^{[00]} & A^{[11]} & A^{[01]} & A^{[10]} 
	\\
	B^{[11]} & B^{[00]} & B^{[10]} & B^{[01]}
	\\
	C^{[01]} & C^{[10]} & C^{[00]} & C^{[11]}
	\\
	D^{[10]} & D^{[01]} & D^{[11]} & D^{[00]}
\end{bmatrix}. 
\]
The explicit form of the representation matrices is given below (we use the same notation as in \eqref{FiniteDalg} for the representation):
\begin{alignat}{2}
	K^0_0 &= \begin{pmatrix} 
	  \sigma_3 & 0 & 0 \\
	  0 & \sigma_3 & 0 \\
	  0 & 0 & 0
	\end{pmatrix}, & \qquad 
	K^+_0 &= \begin{pmatrix}
		\sigma_+ & 0 & 0 \\ 0 & \sigma_+ & 0 \\ 0 & 0 & 0
	\end{pmatrix},
	\notag \\
	K^-_0 &= \begin{pmatrix}
		\sigma_- & 0 & 0 \\ 0 & \sigma_- & 0 \\ 0 & 0 & 0
	\end{pmatrix},
	& L^0_0 &= \begin{pmatrix}
		0 & \sigma_3 & 0 \\ \sigma_3 & 0 & 0 \\ 0 & 0 & 0
	\end{pmatrix},
	\notag \\
	L^+_0 &= \begin{pmatrix}
		0 & \sigma_+ & 0 \\ \sigma_+ & 0 & 0 \\ 0 & 0 & 0 
	\end{pmatrix},
	& 
	L^-_0 &= \begin{pmatrix}
		0 & \sigma_- & 0 \\ \sigma_- & 0 & 0 \\ 0 & 0 & 0
	\end{pmatrix},
	\notag \\
	P^+_0 &= \begin{pmatrix}
		0 & 0 & \sigma_{11} \\ 0 & 0 & i\sigma_+ \\ \sigma_+ & -i\sigma_{22}
	\end{pmatrix},
	& 
	P^-_0 &= \begin{pmatrix}
		0 & 0 & -\sigma_- \\ 0 & 0 & -i\sigma_{22} \\ \sigma_{11} & -i\sigma_- & 0
	\end{pmatrix},
	\notag \\
	Q^+_0 &= \begin{pmatrix}
		0 & 0 & \sigma_+ \\ 0 & 0 & -i\sigma_{11} \\ \sigma_{22} & i \sigma_+ & 0
	\end{pmatrix},
	&
	Q^-_0 &= \begin{pmatrix}
		0 & 0 & -\sigma_{22} \\ 0 & 0 & i\sigma_- \\ \sigma_- & i\sigma_{11} & 0
	\end{pmatrix}
\end{alignat}
where 
\begin{alignat}{2}
	\sigma_3 &= \begin{pmatrix}
		1 & 0 \\ 0 & -1
	\end{pmatrix},
	&\qquad 
	\sigma_+ &= \begin{pmatrix}
		0 & 1 \\ 0 & 0
	\end{pmatrix},
	\qquad 
	\sigma_- = \begin{pmatrix}
		0 & 0 \\ 1 & 0
	\end{pmatrix},
	\notag \\
	\sigma_{11} &= \begin{pmatrix}
		1 & 0 \\ 0 & 0
	\end{pmatrix},
	& \sigma_{22} &= \begin{pmatrix}
		0 & 0 \\ 0 & 1
	\end{pmatrix}.
	\label{6Drepmat}
\end{alignat}
Tensoring with the loop parameter $\lambda$, we obtain a realization of $\g$:
\begin{equation}
	E_m^0 = E_0^0 \otimes \lambda^m, \quad E_m^{\pm} = E_0^{\pm} \otimes \lambda^m, \dots
\end{equation}
Thus the Lax operators with spectral parameter are represented  by $ 6 \times 6$ matrices. 

 Now, we introduce a $ 6 \times 6 $ matrix $\Gamma $ satisfying
\begin{equation}
	\partial_{\mu} \Gamma_{ij} = (\TL_{\mu} \Gamma)_{ij} - \Gamma_{ij} (\TL_{\mu} \Gamma)_{jj}, 
	\quad \mu = t, x
	\label{CondCons}
\end{equation}
where, throughout this subsection, we denote the spacetime coordinates  by $t$ and $x$ in order to emphasize the time independence of the conserved quantities. 
We also note that $\Gamma$ also depends on $\lambda.$ 
With this $\Gamma$, one can obtain the conserved currents
	\begin{equation}
	J_j^t := (\TL_x \Gamma)_{jj}, \qquad J_j^x := -(\TL_t \Gamma)_{jj}. \label{Jdefinition}
\end{equation}
The current conservation is verified as follows:
	\begin{align}
	\partial_t J_j^t &= (\partial_t \TL_x \cdot \Gamma +\TL_x \partial_t \Gamma)_{jj}
	\notag\\
	&\stackrel{\eqref{CondCons}}{=}
	 (\partial_t \TL_x \cdot \Gamma)_{jj} + (\TL_x \TL_t \Gamma)_{jj} - (\TL_x \Gamma)_{jj} (\TL_t \Gamma)_{jj}.
\end{align}
Similarly, we have
	\begin{equation}
	\partial_x J_j^x = -(\partial_x \TL_t \cdot \Gamma)_{jj} - (\TL_t \TL_x \Gamma)_{jj} + (\TL_t \Gamma)_{jj} (\TL_x \Gamma)_{jj}.
\end{equation}
Therefore
	\begin{align}
	\partial_{\mu} J_j^{\mu} = \big( (\partial_t  \TL_x - \partial_x \TL_t + [\TL_x, \TL_t]) \cdot \Gamma \big)_{jj} = 0.
\end{align}

It follows that the following are the constants of motion:
\begin{equation}
	Q_j = \int_{-\infty}^{\infty} dx J_j^t = \int_{-\infty}^{\infty} dx (\TL_x \Gamma)_{jj} \label{chargeformula}
\end{equation}
{\color{mycol}where we assume that all fields vanish at the spatial infinity.}
We introduce the ansatz
\begin{equation}
	\Gamma_{ij} = \sum_{k=0}^{\infty} \frac{\Gamma_{ij}^{(k)}}{\lambda^k}. \label{GammaAnsatz}
\end{equation}
Then, $ Q_j$ can also be expanded in a Laurent series in $\lambda$, which yields an infinite number of conserved quantities. Note that the conserved charges \eqref{chargeformula} are [00]-graded, {\color{mycol}since the currents \eqref{Jdefinition} are defined by the diagonal elements of the matrices. 

Before closing, we comment on the matrix $\Gamma$ and the equation \eqref{CondCons}. 
Recall that one may introduce a linear system of equations whose compatibility condition yields the zero-curvature condition. 
Using a solution $\phi_i$ of such a linear system, one may define $\Gamma_{ij} = \phi_i \phi_j^{-1}$ (see, for example, \cite{Ferreira}). 
In the present case, however, some $\phi_i$ have nontrivial grading, so that their inverses are not well-defined. Therefore, we did not introduce such a linear system; instead, we defined $\Gamma$ via the equation \eqref{CondCons}. 
We also note that the conserved charges are determined only by the Lax operator $\mathcal{L}_x.$ 
}

%%%%%%%%%%%%%%%%%%%%%%%%%%%%%%%%%%%%%%%%%%%%%%%%%%%%%%%%%%%%%%%%%%
%
\section{Hierarchy of $\Z2$-graded integrable equations} \label{SEC:hier}

\subsection{$\Z2$-graded extension of Liouville and sinh-Gordon equations}

We consider the Lax operator consisting of the negative part of $\g$:
\begin{align}
	\mathcal{L}_-&=D_{-\frac{1}{2}}+D_{-1},
	\\
	D_{-\frac{1}{2}} &=\rho_{10}P_0^-+\rho_{01}Q_0^-,
	\nn\\
	D_{-1} &=a_{00}K_{-1}^++b_{00}K_0^-+c_{11}L_{-1}^++d_{11}L_0^-
	\nn
\end{align}
where the coefficients of the basis elements of $\g$ in $D_{-1/2} $ and $ D_{-1}$ depend on $x_{\pm}$. 
The zero-curvature equation \eqref{0curvEq} decomposes into four equations:
\begin{alignat}{2}
	& \text{grade } -1: &    -\partial_+D_{-1}+[A_0,D_{-1}]& =0,
	\nn\\
	& \text{grade } -\frac{1}{2}: & \qquad  -\partial_+D_{-\frac{1}{2}}+[A_0,D_{-\frac{1}{2}}]+[A_{\frac{1}{2}},D_{-1}] &=0,
	\nn \\
	& \text{grade } 0: & \partial_-A_0+[A_{\frac{1}{2}},D_{-\frac{1}{2}}]+[E_1,D_{-1}] & =0,
	\nn \\
	& \text{grade } \frac{1}{2}: & \partial_-A_{\frac{1}{2}}+[E_1,D_{-\frac{1}{2}}] &=0
\end{alignat}
{\color{mycol}where $A_0, A_{\frac{1}{2}}$ and $E_1$ are defined in \eqref{LxDefinition}. }
The grade $-1$ equation is easily integrated, yielding
\begin{align*}
 a_{00}& = k(x_-) e^{2\varphi_{00}} \cosh2\varphi_{11},
 \qquad 
 b_{00}= \ell(x_-)e^{-2\varphi_{00}}\cosh2\varphi_{11},
 \nn \\
 c_{11} & =k(x_-)e^{2\varphi_{00}}\sinh2\varphi_{11},
 \qquad
 d_{11}=- \ell(x_-)e^{-2\varphi_{00}}\sinh2\varphi_{11}
\end{align*}
where $ k(x_-), \ell(x_-)$ are arbitrary [00]-graded functions. 
When integrating over $x_+$, we are free to introduce arbitrary functions of $x_-$. These functions may have nontrivial $\Z2$-gradings, but we keep only the [00]-graded ones {\color{mycol}for the sake of simplicity.} 
The grade $-\frac{1}{2}, 0 $ and $ \frac{1}{2}$ equations yield the equations of motion for $ \rho$'s, $\varphi$'s and $\sigma$'s, respectively:
\begin{align}
	\partial_+\rho_{10} &= -(\partial_+\varphi_{00})\rho_{10}-i(\partial_+\varphi_{11})\rho_{01}+\sigma_{10}b_{00}-i\sigma_{01}d_{11},
	\nn \\
	\partial_+\rho_{01} &= -(\partial_+\varphi_{00})\rho_{01}+i(\partial_+\varphi_{11})\rho_{10}+\sigma_{01}b_{00}+i\sigma_{10}d_{11},
	\nn\\
	\partial_{+-}\varphi_{00} &= a_{00}-b_{00}-\rho_{10}\sigma_{10}-\rho_{01}\sigma_{01},
	\nn \\
	\partial_{+-}\varphi_{11} &= c_{11}- d_{11}+i(\rho_{10}\sigma_{01} - \rho_{01}\sigma_{10}),
	\nn\\
	\partial_-\sigma_{10} &=\rho_{10}, \nn
	\\
	\partial_-\sigma_{01} &=\rho_{01}. \label{sigmaEoM2}
\end{align}
This system of equations involves six dynamical fields, two of which ($ \varphi_{00}, \varphi_{11}$) mutually commute, while the remaining ones ($\rho$'s and $\sigma$'s) are nilpotent. It also contains two arbitrary chiral functions $ k(x_-) $ and $\ell(x_-)$. 
By specifying the arbitrary functions, this system reduces to a $\Z2$-graded extension of well-known equations. 

\bigskip\noindent
a) $ k(x_-) \equiv 0, \ \ell(x_-) \equiv -1: $ $\Z2$-Liouville equation 
\begin{align}
	\partial_{+-}\varphi_{00} &=e^{-2\varphi_{00}}\cosh2\varphi_{11}-\rho_{10}\sigma_{10}-\rho_{01}\sigma_{01},
	\nn\\
	\partial_{+-}\varphi_{11} &= -e^{-2\varphi_{00}}\sinh2\varphi_{11}+i(\rho_{10}\sigma_{01}- \rho_{01}\sigma_{10}),
	\nn \\
	\partial_+\rho_{10}& =-(\partial_+\varphi_{00})\rho_{10}-i(\partial_+\varphi_{11})\rho_{01}
	-e^{-2\varphi_{00}} (\sigma_{10} \cosh 2\varphi_{11} + i \sigma_{01} \sinh 2\varphi_{11}),
	\nn \\
	\partial_+\rho_{01}&=-(\partial_+\varphi_{00})\rho_{01}+i(\partial_+\varphi_{11})\rho_{10}
	-e^{-2\varphi_{00}} (\sigma_{01} \cosh 2\varphi_{11} - i\sigma_{10} \sinh 2\varphi_{11}),
	\label{Z22Liouville}
\end{align}
and the last two equations in \eqref{sigmaEoM2}. 

 Setting all nontrivially graded fields {\color{mycol}([11], [10] and [01]-graded fields) } to zero yields {\color{mycol}the classical ordinary Liouville equation.} 
If we set $\varphi_{11} $ and  the [10]-graded (or [01]-graded) fields to zero, we recover the super-Liouville equation. 

The system of equations \eqref{Z22Liouville} is identical to the $\Z2$-super-Liouville equation obtained by applying the superfield formalism for the finite dimensional $\Z2$-$\osp(1|2)$ \cite{AFITTsuper}. In \cite{AFITTsuper}, an explicit solution in terms of arbitrary chiral functions are constructed by the method of Leznov and Saveliev \cite{LeSa1,LeSa2,LeSa3}. 
The transformations of the coordinates and fields between these two systems are given as follows:
\begin{equation}
  \begin{array}{ccc}
  	\cite{AFITTsuper} & \qquad  & \eqref{Z22Liouville}
  	\\ \hline
  	(x, \bar{x}) & & (-i x_+, -ix_-)
  	\\
  	\varphi_{00} & & -\varphi_{00}
  	\\
  	\varphi_{11} & & -\varphi_{11}
  	\\
  	\psi_{10} & & \sigma_{10}
  	\\
  	\bar{\psi}_{10} & & e^{\varphi_{00}} (\rho_{10} \cosh \varphi_{11} - i\rho_{01} \sinh \varphi_{11})
  	\\
  	\psi_{01} & & -i \sigma_{01}
  	\\
  	\bar{\psi}_{01} & & e^{\varphi_{00}} (\rho_{10} \sinh \varphi_{11} -i \rho_{01} \cosh\varphi_{11} )
  \end{array}
  \label{changeVariables}
\end{equation}
The reason for the coincidence can be understood in the following way. 
Noting that $ a_{00} = c_{11} = 0 $ due to $ k(x_-) = 0$, the present Lax operators $\TL_{\pm} $ contain the following elements of $\g$:
\[
   K^0_0, \ K^{+}_0, \ K^-_1, \ P^{\pm}_0, \ Q^{\pm}_0, \ L^0_0, \ L^{\pm}_0.
\]  
Comparing these with the finite dimensional $\Z2$-$\osp(1|2)$ given in \eqref{FiniteDalg}, the only difference is $E^-_1 $ and $ E_0^-$. Since $[E^-_1, \TL_-] = 0$, $E^-_1$ does not contribute to the equations.  $E_0^-$ does not appear in the superfield Lax operators. Thus, the two approaches yield the same system of equations.

\bigskip\noindent
b)  $ k(x_-) = \ell(x_-) \equiv \frac{1}{2}: $ $\Z2$-sinh-Gordon equation
\begin{align}
	\partial_{+-}\varphi_{00}&= \sinh 2\varphi_{00}\cosh2\varphi_{11}-\rho_{10}\sigma_{10}-\rho_{01}\sigma_{01},
	\nn \\
	\partial_{+-}\varphi_{11}&= \cosh 2\varphi_{00}\sinh2\varphi_{11}+i(\rho_{10}\sigma_{01}-\rho_{01}\sigma_{10}),
	\nn \\
	\partial_+\rho_{10}&= -(\partial_+\varphi_{00})\rho_{10}-i(\partial_+\varphi_{11})\rho_{01} - \frac{1}{2}e^{-2\varphi_{00}} (\sigma_{10} \cosh 2\varphi_{11} + i \sigma_{01} \sinh 2\varphi_{11}),
	\nn \\
	\partial_+\rho_{01}&=-(\partial_+\varphi_{00})\rho_{01}+i(\partial_+\varphi_{11})\rho_{10}- \frac{1}{2}e^{-2\varphi_{00}} (\sigma_{01} \cosh 2\varphi_{11} - i \sigma_{10} \sinh 2\varphi_{11}),
	\label{Z22SG}
\end{align}
and the last two equations in  \eqref{sigmaEoM2}. 

In terms of the  $\bar{\psi}$'s defined in \eqref{changeVariables} and slightly modified $\psi$'s (with the spacetime coordinates and the $\varphi$'s unchanged)
\[
   \psi_{10}= \frac{1}{\sqrt{2}} \sigma_{10}, \qquad \psi_{01} = -\frac{i}{\sqrt{2}}\sigma_{01},
\]
these equations yield the following form:
\begin{align}
	\partial_{+-}\varphi_{00}&= \sinh 2\varphi_{00}\cosh2\varphi_{11} 
	\nn \\
	& + \sqrt{2} e^{-\varphi_{00}} 
	\Big(  (\psi_{10} \bar{\psi}_{10} - \psi_{01} \bar{\psi}_{01}) \cosh\varphi_{11} +  (\psi_{01} \bar{\psi}_{10} - \psi_{10}\bar{\psi}_{01}) \sinh \varphi_{11}\Big),
	\nn \\
	\partial_{+-}\varphi_{11}&= \cosh 2\varphi_{00}\sinh2\varphi_{11}
	\nn \\
	& - \sqrt{2} e^{-\varphi_{00}} 
	\Big( \sinh\varphi_{11} (\psi_{10} \bar{\psi}_{10} - \psi_{01} \bar{\psi}_{01}) + \cosh \varphi_{11} (\psi_{01} \bar{\psi}_{10} - \psi_{10}\bar{\psi}_{01}) \Big),
	\nn \\
	\partial_- \psi_{10} &= \frac{1}{\sqrt{2}} e^{-\varphi_{00}} (\bar{\psi}_{10} \cosh\varphi_{11}  -  \bar{\psi}_{01} \sinh\varphi_{11}),
	\nn \\
	\partial_+ \bar{\psi}_{10} &=- \frac{1}{\sqrt{2}} e^{-\varphi_{00}} (\psi_{10} \cosh\varphi_{11} - \psi_{01} \sinh\varphi_{11}),
	\nn \\
	\partial_- \psi_{01} &= \frac{1}{\sqrt{2}} e^{-\varphi_{00}} (\bar{\psi}_{01} \cosh \varphi_{11} - \bar{\psi}_{10} \sinh\varphi_{11}),
	\nn \\
	\partial_+ \bar{\psi}_{01} &=- \frac{1}{\sqrt{2}} e^{-\varphi_{00}} (\psi_{01} \cosh\varphi_{11} - \psi_{10} \sinh\varphi_{11}).
	\label{Z22SG2}
\end{align}

Setting all nontrivially graded fields to zero yields {\color{mycol}the classical ordinary sinh-Gordon equation.} 
If we set $\varphi_{11} $ and  the [10]-graded (or [01]-graded) fields to zero, the system \eqref{Z22SG2} gives a super extension of the sinh-Gordon equation.

\bigskip\noindent 
c)  $ k(x_-) = -\ell(x_-) \equiv \frac{1}{2}: $ $\Z2$-cosh-Gordon equation
\begin{align}
	\partial_{+-}\varphi_{00} &=\cosh 2\varphi_{00}\cosh2\varphi_{11}-\rho_{10}\sigma_{10}-\rho_{01}\sigma_{01},
	\nn\\
	\partial_{+-}\varphi_{11} &=\sinh 2\varphi_{00} \sinh2\varphi_{11}+i(\rho_{10}\sigma_{01}- \rho_{01}\sigma_{10}),
	\nn \\
	\partial_+\rho_{10}&= -(\partial_+\varphi_{00})\rho_{10}-i(\partial_+\varphi_{11})\rho_{01}   +\frac{1}{2}e^{-2\varphi_{00}} (\sigma_{10} \cosh2\varphi_{11} + i\sigma_{01} \sinh 2\varphi_{11}),
	\nn\\
	\partial_+\rho_{01} &=-(\partial_+\varphi_{00})\rho_{01}+i(\partial_+\varphi_{11})\rho_{10}+\frac{1}{2}e^{-2\varphi_{00}} (\sigma_{01} \cosh2\varphi_{11} - i\sigma_{10} \sinh 2\varphi_{11}),
	\label{Z22CG}
\end{align}
and the last two equations in \eqref{sigmaEoM2}. 

{\color{mycol}
This member in the hierarchy contains two arbitrary functions $k(x_-),\ell(x_-)$. 
By choosing them appropriately, we have seen that $\Z2$-generalization of known integrable systems are obtained from this member. 
}

%%%%%%
\subsection{$\Z2$-graded extension of mKdV equations} \label{SEC:mKdV}

To derive a $\Z2$-graded extension of the mKdV equation, we make the following change:
\[
  x_+ \ \to \ x, \quad x_- \ \to \ t, \quad 
  \partial_+ \varphi_{00} \ \to \ u_{00}, \quad 
  \partial_+ \varphi_{11} \ \to \ u_{11}.
\]
Then $\TL_+$, now denoted $\TL_x$, becomes
\begin{align}
	\mathcal{L}_x = u_{00}K_0^0+u_{11}L_0^0 + \sigma_{10}P_0^++\sigma_{01}Q_0^+ + K_0^++K_1^-
\end{align}
and we take $\TL_-$, now denoted $\TL_t$, as follows:
\begin{align}
	\mathcal{L}_t &=D_{0}+D_{\frac{1}{2}}+D_{1}+D_{\frac{3}{2}}+D_{2}+D_{\frac{5}{2}}+D_{3},
	\nn \\
	 D_0 &=a_{00}K_0^0+b_{11}L_0^0,
	 \nn\\
	 D_{\frac{1}{2}} &=\rho_{10}P_0^++\rho_{01}Q_0^+,
	 \nn\\
	 D_1 &=c_{00}K_0^++d_{00}K_1^-+e_{11}L_0^++f_{11}L_1^-,
	 \nn\\
	 D_{\frac{3}{2}}&=\xi_{10}P_1^-+\xi_{01}Q_1^-,
	 \nn\\
	 D_2 &=h_{00}K_1^0+k_{11}L_1^0,\\
	 D_{\frac{5}{2}} &=\eta_{10}P_1^++\eta_{01}Q_1^+,
	 \nn\\
	 D_3 &=\ell_{00}K_1^++p_{00}K_2^-+r_{11}L_1^++s_{11}L_2^-.
\end{align}
This $\TL_t$ contains a lot of $\Z2$-commutative functions. However, following the procedure described in \S \ref{SEC:formHi}, all of them are determined in terms of  $ u_{00}, u_{11}, \sigma_{10} $ and $ \sigma_{01}. $ 

In this case, the zero-curvature equation decomposes into nine equations of grades $0$ to $4.$  Using the equations of grades $3, 7/2$ and $4$, we obtain $ p_{00} = \ell_{00} = p(x_-)$ and $ s_{11} = r_{11} = 0. $ Here, we retain only the [00]-graded arbitrary function upon integration over $x$. Now, we set $ p(x_-) = 4. $ Then, after a lengthy computation involving the equations of grades $1$ through $5/2$, we obtain the expression for the $\Z2$-graded functions appearing in $\TL_t$. The [00] and [11]-graded functions are given by
\begin{align}
  a_{00} &= u_{00}'' -2u_{00}^3 -6 u_{00} u_{11}^2 + 3 \Sigma_{00}' + 6u_{00} \Sigma_{00} + 6i u_{11} \Sigma_{11},
  \notag \\
  b_{11} &= u_{11}'' -2u_{11}^3 - 6 u_{00}^2 u_{11} + 3i \Sigma_{11}' + 6i u_{00} \Sigma_{11} + 6 u_{11} \Sigma_{00},
  \notag\\
  c_{00} &=-2( u_{00}' + u_{00}^2 + u_{11}^2 - \Sigma_{00}),
  \notag \\
  d_{00} &= 2 u_{00}' - 2 (u_{00}^2 + u_{11}^2) + 6 \Sigma_{00},
  \notag \\
  e_{11} &= -2 u_{11}' - 4 u_{00} u_{11} + 2i \Sigma_{11},
  \notag \\
  f_{11} &= 2u_{11}' - 4u_{00} u_{11} + 6i \Sigma_{11},
  \notag\\
  h_{00} &= 4 u_{00}, \qquad k_{11} = 4 u_{11} \notag
\end{align}
where the prime denotes the $x$-derivative, i.e., $ u' := \partial_x u $ and
\[
	\Sigma_{00} =\sigma_{10}' \sigma_{10}+ \sigma_{01}' \sigma_{01},
	\qquad 
	\Sigma_{11} = \sigma_{01}' \sigma_{10}- \sigma_{10}' \sigma_{01}.
\]
The [10] and [01]-graded functions are given by
\begin{align}
	\rho_{10} &= 4(\sigma_{10}'' + u_{00} \sigma_{10}' + i u_{11} \sigma_{01}') +  d_{00}\sigma_{10} + if_{11} \sigma_{01},
	\notag \\
	\rho_{01} &= 4(\sigma_{01}'' +   u_{00} \sigma_{01}' - i u_{11} \sigma_{10}') +  d_{00}\sigma_{01} - if_{11} \sigma_{10},
	\notag \\
	\xi_{10} &= -4\sigma_{10}', \qquad \xi_{01} = -4\sigma_{01}',
	\notag\\
	\eta_{10} &= 4\sigma_{10},  \qquad \quad \eta_{01} = 4\sigma_{01}. \notag
\end{align}
With these results, we can obtain the equations of motion from the grade $0$ and the grade $1/2$ equations:
\begin{align}
	\partial_t u_{00} &=  a_{00}', \qquad \partial_t u_{11} = b_{11}',
	\nn\\
	\partial_t \sigma_{10} &= 4 \sigma_{10}''' - 6 U_{00} \sigma_{10}' - 6i U_{11} \sigma_{01}' - 3 U_{00}' \sigma_{10} - 3i U_{11}' \sigma_{01},
	\notag \\
	\partial_t \sigma_{01} &= 4 \sigma_{01}''' -6 U_{00} \sigma_{01}' +6i U_{11} \sigma_{10}' - 3 U_{00}' \sigma_{01} +3i U_{11}' \sigma_{10} 
	\label{Z22mKdV}
\end{align}
where
\begin{equation}
   U_{00} := -u_{00}' + u_{00}^2 + u_{11}^2, 
  \qquad 
  U_{11} := -u_{11}' + 2u_{00} u_{11}. \label{Miura}
\end{equation}

 If we set all the nontrivially graded variables to zero, we recover the mKdV equation of the form
\[
  \partial_t u_{00} = u_{00}''' -6 u_{00}' u_{00}^2.
\]
Therefore, the system of equations \eqref{Z22mKdV} can be regarded as a $\Z2$-graded extension of the mKdV equation. If we set $ u_{11} = \sigma_{01} = 0, $ then \eqref{Z22mKdV} gives a super-extension of the mKdV equation:
\begin{align}
	\partial_t u_{00} &= u_{00}''' -6 u_{00}' u_{00}^2 + 3 (\sigma_{10}''\sigma_{10} + 2u_{00} \sigma_{10}'\sigma_{10})',
	\nn\\
	\partial_t \sigma_{10} &= 4\sigma_{10}''' - 6(-u_{00}'+u_{00}^2)\sigma_{10}'  - 3 (-u_{00}'' +2 u_{00}' u_{00}) \sigma_{10}.
\end{align}

%%%%%%%%%%%%%%%%%%%%%%%%%%%%%%%%%%%%%%%%%%%%%%%%%%%%%%%%%%%%%%%%%%%%%%%%%%%%%%%%%%%
%
\section{$\Z2$-KdV equation and conserved quantities}
\setcounter{equation}{0}

\subsection{Miura transformation and  $\Z2$-KdV equation} \label{SEC:MiuraCon}

A remarkable property of the $\Z2$-mKdV equation derived in \S \ref{SEC:mKdV} is that the functions $ U_{00}$ and $ U_{11}$, which appear in the equations for $\sigma_{10}$ and $ \sigma_{01}$, provide the Miura transformation. 
The corresponding $\Z2$-graded version of KdV equation is given by
\begin{align}
	\partial_t U_{00} &= A_{00}' + 6 (U_{00} \Sigma_{00}' + i U_{11} \Sigma_{11}'),
	\nn \\
	 \partial_t U_{11} &= B_{11}' + 6(U_{11} \Sigma_{00}' + i U_{00} \Sigma_{11}') 
	 \label{Z22KdV}
\end{align}
where
\begin{align}
	A_{00} &= U_{00}'' - 3(U_{00}^2+ U_{11}^2+\Sigma_{00}'') + 6(U_{00} \Sigma_{00} + i U_{11} \Sigma_{11}),
	\nn \\
	B_{11} &= U_{11}'' -3 (2 U_{00} U_{11}+ i\Sigma_{11}'') + 6( U_{11} \Sigma_{00} + i U_{00} \Sigma_{11}).
\end{align}
The equations for $\sigma_{10}$ and $ \sigma_{01}$ remain unchanged under this transformation. 
This can be verified by the following computation:
\begin{align}
	(-\partial_x + 2 u_{00}) (-\partial_t u_{00} + a_{00}') &+ 2 u_{11} (-\partial_t u_{11} + b_{11}')
	\nn \\
	&= -\partial_t U_{00} + A_{00}' + 6 (U_{00} \Sigma_{00}' + i U_{11} \Sigma_{11}'), 
	\nn \\[3pt]
	(-\partial_x + 2 u_{00}) (-\partial_t u_{11} + b_{11}') &+ 2 u_{11} (-\partial_t u_{00} + a_{00}')
	\nn\\
	&= -\partial_t U_{11} + B_{11}' + 6(U_{11} \Sigma_{00}' + i U_{00} \Sigma_{11}').
\end{align}
It is easy to see that these equations reduce to the KdV equation by setting $ U_{11} = \sigma_{10} = \sigma_{01} = 0.$ If we keep $\sigma_{10}, $ then they reduce to the super-KdV equation:
\begin{align}
	\partial_t U_{00} &= U_{00}''' -6U_{00} U_{00}' + 6U_{00}'\sigma_{10}' \sigma_{10} + 12 U_{00} (\sigma_{10}' \sigma_{10})' - 3(\sigma_{10}' \sigma_{10})''',
	\nn\\
	\partial_t \sigma_{10} &= 4\sigma_{10}''' - 6U_{00}\sigma_{10}'  - 3 U_{00}' \sigma_{10}.
\end{align}

We remark that the Miura transformation \eqref{Miura} can be recast into the form of matrix Riccati equation. Introducing the matrices
\begin{equation}
	U = \begin{pmatrix}
		U_{00} & U_{11} \\ U_{11} & U_{00}
	\end{pmatrix},
	\quad
	Y = -\begin{pmatrix}
		u_{00} & u_{11} \\ u_{11} & u_{00}
	\end{pmatrix},
\end{equation}
the Miura transformation then takes the following form:
\begin{equation}
	U = Y' + Y^2.
\end{equation}

As in the case of the KdV and mKdV equations, the existence of a Miura transformation implies a gauge equivalence between the Lax operators of the corresponding $\Z2$-graded equations. 
The gauge transformation in the $\Z2$-graded case is given by
\begin{align}
	\mathcal{L}_{\mu} &\to g^{-1} \mathcal{L}_{\mu} g - g^{-1}\partial_{\mu} g, \quad \mu = t, x
	\nn \\
	g &= \exp(u_{00}K_{-1}^+ + u_{11} L_{-1}^+).
\end{align}
This yields the Lax operator for the $\Z2$-KdV equation \eqref{Z22KdV}, which are explicitly given by
\begin{align}
	\mathcal{L}_x &= U_{00}K_{-1}^+ + U_{11} L_{-1}^+ + \sigma_{10} P_0^+ + \sigma_{01} Q_0^+ + K_0^+ + K_1^-,
	\nn\\
	\mathcal{L}_t &= D_{-1} + D_0 + D_{\frac{1}{2}} + D_1 + D_{\frac{3}{2}} + D_2 + D_{\frac{5}{2}} + D_3
\end{align}
where
\begin{align}
	D_{-1} &=  (A_{00} + U_{00}^2 + U_{11}^2 ) K_{-1}^+ + (B_{11}+ 2U_{00} U_{11} ) L_{-1}^+,
	\nn\\
	D_0 &= \frac{1}{2} (d_{00}' K_0^0 + f_{11}' L_0^0),
	\nn\\
	D_{\frac{1}{2}} &= (4\sigma_{10}''+ d_{00} \sigma_{10} + if_{11}\sigma_{01}) P_0^+ 
	+ (4\sigma_{01}'' + d_{00} \sigma_{01} -if_{11}\sigma_{10}) Q_0^+,
	\nn \\
	D_1 &= 2(U_{00}+ \Sigma_{00}) K_0^+ + 2(U_{11}+ i\Sigma_{11}) L_0^+ + d_{00} K_1^- + f_{11} L_1^-,
	\nn\\
	D_{\frac{3}{2}} &= -4(\sigma_{10}' P_1^- + \sigma_{01}' Q_1^-),
	\nn\\
	D_{\frac{5}{2}} &= 4(\sigma_{10} P_1^+ + \sigma_{01} Q_1^+),
	\nn\\
	D_3 &= 4(K_1^+ + K_2^-)
\end{align}
with
\[
  d_{00} = 2(-U_{00} + 3\Sigma_{00}), \qquad
  f_{11} = 2(-U_{11}+ 3i\Sigma_{11}).
\]

%%%%%%%%%%%%%%%%%%%%%%%%%%%%%%%%%%%%%%%%%%
\subsection{Conserved charges for $\Z2$-KdV and $\Z2$-mKdV equations} \label{SEC:ChargKdV}

We demonstrate the computation of conserved charges described in \S \ref{SEC:Charges}. 
To this end, {\color{mycol}we switch from the principal gradation \eqref{PrincipleG} to the homogeneous gradation \eqref{HomoG}, since the computation of conserved charges becomes much easier. }
This does not alter the resulting $\Z2$-KdV equation, since there is an injective homomorphism between the two gradations $ f : \g \to \g $ defined by
\begin{alignat}{4}
	&\g_{2m} & \ : \ &\langle \ K_m^0,\; L_m^0 \ \rangle & 
	& \xrightarrow{\quad f \quad} & \quad 
	& \langle \ K^0_{4m},\; L^0_{4m}  \ \rangle
	\nn \\
	&\g_{2m\pm\frac{1}{2}}  & \ : \ &\langle \ P_m^\pm,\; Q_m^\pm \ \rangle & 
	& \xrightarrow{\quad f \quad} & 
	&\langle \ P^{\pm}_{4m\pm 1},\; Q^{\pm}_{4m\pm 1}  \ \rangle
	\nn \\
	&\g_{2m+1}  & \ : \ &\langle \ K_m^+,\; K_{m+1}^-,\; L_m^+,\; L_{m+1}^- \ \rangle & 
	& \xrightarrow{\quad f \quad} & 
	&\langle \ K^+_{4m+2},\; K^-_{4m+2},\; L^+_{4m+2},\; L^-_{4m+2}  \ \rangle \notag
\end{alignat}
In the homogeneous gradation, the Lax operator takes the form
\begin{equation}
	\mathcal{L}_x = U_{00}K_{-2}^+ + U_{11} L_{-2}^+ + \sigma_{10} P_1^+ + \sigma_{01} Q_1^+ + K_2^+ + K_2^-.
\end{equation}
Its matrix representation in the six-dimensional module introduced in
\S \ref{SEC:Charges} is given by the following form:
\begin{equation}
	\mathcal{L}_x = 
	\begin{pmatrix}
		0 & \lambda^2 + \lambda^{-2} U_{00} & 0 & \lambda^{-2} U_{11} & \lambda \sigma_{10} & \lambda \sigma_{01}
		\\
		\lambda^2 & 0 & 0 & 0 & 0 & 0
		\\
		0 & \lambda^{-2} U_{11} & 0 & \lambda^2 + \lambda^{-2} U_{00} & i\lambda \sigma_{01} & -i\lambda \sigma_{10}
		\\
		0 & 0 & \lambda^2 & 0 & 0 & 0
		\\
		0 & -\lambda \sigma_{10} & 0 & i\lambda \sigma_{01} & 0 & 0
		\\
		0 & -\lambda \sigma_{01} & 0 & -i\lambda \sigma_{10} & 0 & 0
	\end{pmatrix}.
\end{equation}

We consider the conserved charges derived from the current
\[
J_2^t = (\mathcal{L}_x \Gamma)_{22} =  \lambda^2 \Gamma_{12}.
\]  
To compute them, we need to determine  $ \Gamma_{ij} $ by  solving the equations
\begin{equation}
	\partial_x \Gamma_{i2} = (\mathcal{L}_x \Gamma)_{i2} - \Gamma_{i2} J^t_2, \quad i = 1, 2, \dots, 6
\end{equation}
The $i =2$ equation $ \partial_x \Gamma_{22} = (1- \Gamma_{22}) J^t_2$ is easily solved, yielding  $ \Gamma_{22} = 1. $ 
To solve the remaining equations
\begin{alignat}{2}
	& i = 1 & \ : \ \partial_x \Gamma_{12} &= \lambda^2 + \lambda^{-2} U_{00} + \lambda^{-2} U_{11} \Gamma_{42} + \lambda \sigma_{10} \Gamma_{52} + \lambda \sigma_{01} \Gamma_{62} -\lambda^2 \Gamma_{12}^2,
	\nn\\[3pt]
	& i = 3 & \ : \ \partial_x \Gamma_{32} &= \lambda^{-2} U_{11} + (\lambda^2 + \lambda^{-2} U_{00}) \Gamma_{42} + i\lambda \sigma_{01} \Gamma_{52} - i\lambda \sigma_{10} \Gamma_{62} - \lambda^2 \Gamma_{12} \Gamma_{32},
	\nn \\[3pt]
	& i = 4 & \ : \ \partial_x \Gamma_{42} &= \lambda^2 \Gamma_{32} - \lambda^2 \Gamma_{12} \Gamma_{42},
	\nn \\[3pt]
	& i = 5 & \ : \ \partial_x \Gamma_{52} &= -\lambda \sigma_{10} + i\lambda \sigma_{01} \Gamma_{42} - \lambda^2 \Gamma_{12} \Gamma_{52}, 
	\nn \\[3pt]
	& i = 6 & \ : \ \partial_x \Gamma_{62} &= -\lambda \sigma_{01} - i\lambda \sigma_{10} \Gamma_{42} - \lambda^2 \Gamma_{12} \Gamma_{62}, 
\end{alignat}
it is convenient to introduce an ansatz that is slightly different from \eqref{GammaAnsatz}:
\begin{equation}
	\Gamma_{i2} = 
	\begin{cases}
		\displaystyle \sum_{k \geq 0} \frac{\Gamma_{i2}^{(2k)}}{\lambda^{2k}} & i = 1, 3, 4
		\\[18pt]
		\displaystyle \sum_{k \geq 0} \frac{\Gamma_{i2}^{(2k+1)}}{\lambda^{2k+1}} & i = 5, 6 
	\end{cases}
\end{equation}
To obtain the expansion of $ J_2^t$ in $\lambda,$ we need to determine $ \Gamma_{12}^{(2k)}$ order by order. 

By examining each order in $\lambda,$ we can determine $ \Gamma_{i2}^{(k)}.$ 
\begin{alignat*}{3}
	& \lambda^2 & \ : \  \Gamma_{12}^{(0)} &= \epsilon = \pm 1, & \qquad \Gamma_{42}^{(0)} &= \epsilon \Gamma_{32}^{(0)},
	\\
	& \lambda^1 & \ : \  \Gamma_{52}^{(1)} &= -\epsilon \sigma_{10} + i\sigma_{01} \Gamma_{32}^{(0)},
	& 
	\Gamma_{62}^{(1)} &= -\epsilon \sigma_{01} -i \sigma_{10} \Gamma_{32}^{(0)},
	\\
	&\lambda^0 & \ : \  \Gamma_{12}^{(2)} &= 0, & \partial_x \Gamma_{32}^{(0)} &= 0, \qquad \Gamma_{42}^{(2)} = \epsilon \Gamma_{32}^{(2)}.
\end{alignat*}
We take $ \Gamma_{32}^{(0)} = 0.$ Proceeding further, we  obtain
\begin{alignat}{3}
	&\lambda^{-1} & \ : \  \Gamma_{52}^{(3)} &= i\sigma_{01} \Gamma_{32}^{(2)} + \sigma_{10}', 
	& \qquad 
	\Gamma_{62}^{(3)} &= -i\sigma_{10} \Gamma_{32}^{(2)} + \sigma_{01}',
	\nn\\
	& \lambda^{-2} & \ : \ 2\epsilon \Gamma_{12}^{(4)} &= U_{00} - \Sigma_{00}, 
	\nn\\
	& & 
	2\partial_x \Gamma_{32}^{(2)} &= U_{11} - i \Sigma_{11}, 
	& \Gamma_{42}^{(4)} &= \epsilon \Gamma_{32}^{(4)} - \partial_x \Gamma_{32}^{(2)}, 
	\label{G42G32}\\
	& \lambda^{-3} & \ : \ \epsilon \Gamma_{52}^{(5)} &= i\sigma_{01} \Gamma_{42}^{(4)}  + \epsilon \Gamma_{12}^{(4)} \sigma_{10} -\partial_x \Gamma_{52}^{(3)},
	\nn \\
	& & \epsilon \Gamma_{62}^{(5)} &= -i\sigma_{10} \Gamma_{42}^{(4)} + \epsilon \Gamma_{12}^{(4)} \sigma_{01} -\partial_x \Gamma_{62}^{(3)}.
	\nn
\end{alignat}
The computation of the subsequent orders is more involved. Using the relations obtained thus far, we find the following:
\begin{alignat}{2}
	&\lambda^{-4} & \ : \ &\Gamma_{12}^{(6)} = \frac{1}{2}\partial_x\big( (\Gamma_{32}^{(2)})^2 + \Sigma_{00} - \epsilon \Gamma_{12}^{(4)} \big),
	\nn\\
	& & & \partial_x (\Gamma_{42}^{(4)} + \epsilon \Gamma_{32}^{(4)}) = i\Sigma_{11}',  
	\label{G42}\\
	& & & \partial_x (\Gamma_{42}^{(4)} - \epsilon \Gamma_{32}^{(4)}) = -2\epsilon( \Gamma_{42}^{(6)} - \epsilon \Gamma_{32}^{(6)} ) - 2 \epsilon \Gamma_{12}^{(4)} \Gamma_{32}^{(2)} +i\Sigma_{11}'.
	\nn
\end{alignat}
Combining \eqref{G42G32} with the relation obtained from integrating \eqref{G42}, we obtain 
\begin{equation}
	\Gamma_{42}^{(4)} = \frac{1}{2}(i\Sigma_{11}-\partial_x \Gamma_{32}^{(2)}).
\end{equation}
The next order gives two relations:
\begin{alignat*}{2}
	&\lambda^{-5} & \ : \ 
	\epsilon \Gamma_{52}^{(7)} &= i\sigma_{01} \Gamma_{42}^{(6)} - \Gamma_{12}^{(4)} \Gamma_{52}^{(3)} + \epsilon \Gamma_{12}^{(6)} \sigma_{10} - \partial_x \Gamma_{52}^{(5)},
	\\
	& & \epsilon \Gamma_{62}^{(7)} &= -i\sigma_{10} \Gamma_{42}^{(6)} - \Gamma_{12}^{(4)} \Gamma_{62}^{(3)} + \epsilon \Gamma_{12}^{(6)} \sigma_{01} - \partial_x \Gamma_{62}^{(5)}.
\end{alignat*}
Taking into account the relations obtained thus far, we can determine $\Gamma_{12}^{(8)}$ from the computation at the next order:
\begin{align}
	2\epsilon \Gamma_{12}^{(8)} &= -\frac{1}{4}(U_{00}^2 + U_{11}^2) + \frac{3}{2}(U_{00} \Sigma_{00} + i U_{11} \Sigma_{11}) + \sigma_{10}'' \sigma_{10}' + \sigma_{01}'' \sigma_{01}'
	\nn \\
	&-(\Gamma_{12}^{(6)} -i\Sigma_{11} \Gamma_{32}^{(2)}+ \Sigma_{00}')'.
\end{align}
One may repeat this computation to higher order in $\lambda$. 

The conserved charge also admits an expansion in $\lambda$:
\begin{equation}
	Q_2 = \int dx \lambda^2 \Gamma_{12} = \sum_{k \geq 0} \int dx \frac{\Gamma_{12}^{(2k)}}{\lambda^{2k-2}} \equiv \sum_{k \geq 0} Q_2^{(2k)}.
\end{equation}
The nontrivial conserved charges for the $\Z2$-KdV equation are given  (up to an overall constant factor) by
\begin{align}
	Q_2^{(4)} &= \int dx (U_{00} - \Sigma_{00}),
	\nn \\
	Q_2^{(8)} &= \int dx \big( U_{00}^2+U_{11}^2 -6(U_{00} \Sigma_{00} + i U_{11}\Sigma_{11}) -4(\sigma_{10}'' \sigma_{10}' + \sigma_{01}'' \sigma_{01}') \big),
	\nn \\
	& \cdots  
\end{align}
A direct computation confirms the conservation of these quantities.

These conserved charges are mapped to those of the $\Z2$-mKdV equation through the Miura transformation \eqref{Miura}. They are given by
\begin{align}
	Q_2^{(4)} &\to \int dx (u_{00}^2 + u_{11}^2 -\Sigma_{00}),
	\nn \\
	Q_2^{(8)} &\to \int dx \big( u_{00}^4 + u_{11}^4 + 6u_{00}^2 u_{11}^2 + (u_{00}')^2 + (u_{11}')^2 
	\nn \\
	 & \qquad + 6(u_{00}' - u_{00}^2 - u_{11}^2) \Sigma_{00} + 6i(u_{11}' - 2u_{00}u_{11}) \Sigma_{11} -4(\sigma_{10}'' \sigma_{10}' + \sigma_{01}'' \sigma_{01}')
	 \big). 
\end{align}
In addition, it follows from the $\Z2$-mKdV equation \eqref{Z22mKdV} that the following quantity of  [11]-grading is conserved:
\begin{equation}
	\int dx u_{11}. \label{11conserved}
\end{equation}
Thus, the $\Z2$-mKdV equation has at least one conserved charge with nontrivial grading. 
Finally, we comment on the [11]-graded conserved charge for $\Z2$-KdV equation corresponding to \eqref{11conserved}.  
A natural candidate is the integral of $  (U_{11} -i \Sigma_{11}).$ Indeed, one may verify its conservation by using the equations of motion. However, from \eqref{G42G32} it follows that this  integration vanishes. Therefore, there is no counterpart of \eqref{11conserved} for $\Z2$-KdV equation.

%%%%%%%%%%%%%%%%%%%%%%%%%%%%%%%%%%%%%%%%%%%%%%%%%%%%%%%%%%%%%%%%%%%%%%%%%
\section{Conclusion} \label{SEC:Conclusion}

We have derived a $\Z2$-graded integrable hierarchy that includes $\Z2$-graded extensions of the Liouville, sinh-Gordon, cosh-Gordon, and mKdV equations. We have shown that the Miura transformation can be directly read off from the $\Z2$-mKdV equation and used it to define the corresponding $\Z2$-KdV equation. The Lax operators for the $\Z2$-KdV equation were derived via a gauge transformation associated with the Miura transformation. 
We also presented explicit formulas for conserved charges of the $\Z2$-KdV and $\Z2$-mKdV equations. Moreover, the analysis reveals the existence of conserved charges with [11]-grading. 

From the viewpoint of mathematical physics, this work elucidates how standard constructions, such as Miura transformations and conserved quantities, are modified in the $\Z2$-graded setting, thereby extending the algebraic theory of integrable systems. 
{\color{mycol}An interesting possible direction for future investigation is the construction of the classical $r$-matrix associated with the $\Z2$-$\osp(1|2)$. 
Another interesting direction is the study of the so-called B\"acklund-gauge transformations which connect members of different hierarchies \cite{GomesTypeII,GomesMiura,GomesGauge}.  
It is natural to expect that such transformations can be constructed in the $\Z2$-setting, given the parallelism between  the formalism of the present paper and that of previous works on affine Lie algebras. Such transformations are expected to yield new integrable systems associated with  $\Z2$-graded Lie superalgebras.

We have focused on $\Z2$-graded extensions of Lie superalgebras, in particular on $\Z2$-$\osp(1|2)$, in order to derive a $\Z2$-graded extension of KdV-type equations. 
It is natural to extend the present construction to higher-rank $\Z2$-graded Lie superalgebras. To carry out such an extension, several issues must be addressed. 
First, we need to define an appropriate $\Z2$-graded Lie superalgebras, since $\Z2$-graded extension of a given Lie (super)algebra is not unique.  
Recall that a $\Z2$-graded extended algebra has a larger dimension than the original Lie (super)algebra. Therefore, computations based solely on the defining relations, as in the present work, are not efficient. It is essential to understand the structure of the extended algebra (such as its Cartan subalgebra and root space decomposition) and to make effective use of this structural information in computations. 
We also need to determine the irreducible representations of the extended algebra. 
It is therefore desirable to further develop the mathematical aspects of $\Z2$-graded Lie superalgebras in order to enable the systematic construction of a broad class of new integrable systems characterized by $\Z2$-graded Lie superalgebraic structures.
}

We also remark that $\Z2$-graded Lie superalgebras constitute only a special class of color Lie (super)algebras, in which Lie (super)algebras are generalized by arbitrary Abelian groups \cite{scheu}. The general structure of color Lie algebras of $\mathfrak{gl}$ type has been studied in detail \cite{zhang2025}. By defining color extensions of $\mathfrak{so}$, $\mathfrak{sp}$, or $\osp$ types in an appropriate manner, one may consider integrable systems associated with these color algebras. 
This will be an interesting future work.

%%%%%%%%%%%%%%%%%%%%%%%%%%%%%%%%%%%%%%%%%%%%%%%%%%%%%%%%%%%%%%%%%%%%
\section*{Acknowledgements}

The authors would like to thank F. Toppan and Z. Kuznetsova for valuable discussions. 
N. A. is supported by JSPS KAKENHI Grant Number JP23K03217. 
R. I. is supported by JST SPRING, Grant Number JPMJSP2139.

%%%%%%%%%%%%%%%%%%%%%%%%%%%%%%%%%%%%%%%%%%%%%%%
\bibliographystyle{JHEP} 
\bibliography{Integrable}

@article{rw1,
  author  = {Rittenberg, V. and Wyler, D.},
  title   = {Generalized superalgebras},
  journal = {Nuclear Physics B},
  volume  = {139},
  number  = {3},
  year    = {1978},
  pages   = {189--202},
  doi     = {10.1016/0550-3213(78)90186-4},
  issn    = {0550-3213}
}

@article{rw2,
    author = {V. Rittenberg and D. Wyler},
    title = "{Sequences of $\bm{Z}_2 \otimes \bm{Z}_2$ graded {L}ie algebras and superalgebras}",
    journal = {J. Math. Phys.},
    volume = {19},
    pages = {2193--2200},
    year = {1978},
    month = {10},
    doi = {10.1063/1.523552},
    number = {10},
    issn = {0022-2488},
    url = {https://doi.org/10.1063/1.523552},
}

@article{scheu,
    author = {M. Scheunert},
    title = "{Generalized {L}ie algebras}",
    journal = {J. Math. Phys.},
    volume = {20},
    pages = {712--720},
    year = {1979},
    month = {04},
    doi = {10.1063/1.524113},
    url = {https://doi.org/10.1063/1.524113},
    number = {4},
    issn = {0022-2488},
}

@article{zhang2025,
  title={Invariants and representations of the {$\Gamma$}-graded general linear {Lie} $\omega $-algebras},
  author={Zhang, R. B.},
  year={2025},
  journal={},
  eprint={arXiv:2509.21795[math.RT]}
}

@article{AmaAi,
    author = {K. Amakawa and N. Aizawa},
    title = {A classification of lowest weight irreducible modules over $\mathbb{Z}_2^2$-graded extension of $osp(1|2)$},
    journal = {J. Math. Phys.},
    volume = {62},
    pages = {043502},
    year = {2021},
    month = {04},
    doi = {10.1063/5.0037493},
    number = {4},
    issn = {0022-2488},
    url = {https://doi.org/10.1063/5.0037493},
}

@article{AiSe,
   author = "N. Aizawa and J. Segar",
   title  = "{Affine extensions of $\mathbb{Z}_2^2$-graded $osp(1|2)$  and {V}irasoro algebra}",
  journal={Int. J. Geom. Methods Mod. Phys.},
  pages={2540052},
   volume={23}, 
   year ={2026},
   doi = {10.1142/S0219887825400523},
}

@article{Aizawa_2020,
author = {N. Aizawa and K. Amakawa and S. Doi},
year = {2020},
month = {jan},
publisher = {IOP Publishing},
volume = {53},
pages = {065205},
title = {{$\mathcal N$}-extension of double-graded supersymmetric and superconformal quantum mechanics},
journal = {J. Phys. A: Math. Theor.},
doi = {10.1088/1751-8121/ab661c},
number = {6},
url = {https://doi.org/10.1088/1751-8121/ab661c},
}

@article{LeSa1,
   author  = "A. N. Leznov and M. V. Saveliev",
   title   = "{Representation of zero curvature for the system of nonlinear partial differential equations $x_{\alpha,z \bar{z}} = \exp(kx)_{\alpha}$ and its integrability}",
   journal = "Lett. Math. Phys.",
   volume  = "3",
   pages   = "489--494",
   year    = "1979",
   doi = "10.1007/BF00401930",
}

@article{LeSa2,
   author  = "A. N. Leznov and M. V. Saveliev",
   title   = "Exactly and completely integrable nonlinear dynamical systems",
   journal = "Acta. Appl. Math.",
   volume  = "16",
   pages   = "1-74",
   year    = "1989",
   doi = "10.1007/BF00046886",
}

@article{LeSa3,
   author  = "A. N. Leznov and M. V. Saveliev",
   title   = "Two-dimensional exactly and completely integrable dynamical systems", 
   journal = "Comm. Math. Phys.",  
   volume  = "89",
   pages   = "59-75",
   year    = "1983",
   doi = {https://doi.org/10.1007/BF01219526},
}

@article{babelon1990conformal,
  title="{Conformal affine $sl_2$ {Toda} field theory}",
  author={O. Babelon and L. Bonora},
  journal={Phys. Lett. B},
  volume={244},
  pages={220--226},
  year={1990},
  doi = {10.1016/0370-2693(90)90059-F},
  publisher={Elsevier},
  number={2},
}

@article{toppan1991superliouville,
  title={On the {superLiouville} theory},
  author={F. Toppan},
  journal={Phys. Lett. B},
  volume={260},
  pages={346--352},
  year={1991},
  publisher={Elsevier},
  number={3-4},
  doi={10.1016/0370-2693(91)91623-4},
}

@article{toppanZhan1992,
  title="{Superconformal affine {Liouville} theory}",
  author={F. Toppan and Y.-Z. Zhang},
  journal={Phys. Lett. B},
  volume={292},
  pages={67--76},
  year={1992},
  publisher={Elsevier},
  number={1-2},
  doi={10.1016/0370-2693(92)90609-8},
}

@article{aratyn2004algebraic,
  title={Algebraic construction of integrable and super integrable hierarchies},
  author={H. Aratyn and J. F. Gomes and A. H. Zimerman},
  eprint={hep-th/0408231},
  year={2004}
}

@Inbook{Aratyn2001,
author="H. Aratyn
and J. F. Gomes  
and A. H. Zimerman  
and E. Nissimov 
and S. Pacheva",
editor="H. Aratyn
and A. S. Sorin",
title="Symmetry flows, conservation laws and dressing approach to the integrable models",
bookTitle="Integrable Hierarchies and Modern Physical Theories",
year="2001",
publisher="Springer Netherlands",
address="Dordrecht",
pages="243--275",
doi="10.1007/978-94-010-0720-7_8",
isbn="978-94-010-0720-7",
url="https://doi.org/10.1007/978-94-010-0720-7_8"
}

@article{GomesTypeII,
doi = {10.1088/1751-8113/48/40/405203},
url = {https://doi.org/10.1088/1751-8113/48/40/405203},
year = {2015},
volume = {48},
number = {40},
pages = {405203},
author = {Gomes, J F and Retore, A L and Zimerman, A H},
title = {Construction of type-{II} {B}\"acklund transformation for the {mKdV} hierarchy},
journal = {J.Phys. A:Math. Theor.},
}

@article{GomesMiura,
doi = {10.1088/1751-8113/49/50/504003},
url = {https://doi.org/10.1088/1751-8113/49/50/504003},
year = {2016},
publisher = {IOP Publishing},
volume = {49},
number = {50},
pages = {504003},
author = {Gomes, J F and Retore, A L and Zimerman, A H},
title = {Miura and generalized {B}\"acklund transformation for {KdV} hierarchy},
journal = {J.Phys. A:Math. Theor.},
}

@article{GomesGauge,
doi = {10.1088/1751-8121/ac2718},
url = {https://doi.org/10.1088/1751-8121/ac2718},
year = {2021},
publisher = {IOP Publishing},
volume = {54},
number = {43},
pages = {435201},
author = {de Carvalho Ferreira, J M and Gomes, J F and Lobo, G V and Zimerman, A H},
title = {Gauge {M}iura and {B}\"acklund transformations for generalized {$A_n$}-{KdV} hierarchies},
journal = {J.Phys. A:Math. Theor.},
}

@article{MixedHi,
author = {J. F. Gomes and G. R. de Melo  and L. H. Ymai and A. H. Zimerman},
title = "{Nonautonomous mixed {mKdV}-sinh-{Gordon} hierarchy}",
journal = {J. Phys. A: Math. and Theor.},
year = {2010},
month = {aug},
publisher = {},
volume = {43},
pages = {395203},
doi = {10.1088/1751-8113/43/39/395203},
number = {39},
url = {https://doi.org/10.1088/1751-8113/43/39/395203},
}

@article{BruSG,
author = {A. J. Bruce},
title = {Is the $\mathbb{Z}_2 \times \mathbb{Z}_2$-graded sine-{G}ordon equation integrable ?},
journal = {Nucl. Phys. B},
volume = {971},
pages = {115514},
year = {2021},
doi = {10.1016/j.nuclphysb.2021.115514},
issn = {0550-3213},
url = {https://www.sciencedirect.com/science/article/pii/S055032132100211X},
}

@article{AIKTTslint,
author = {N. Aizawa and R. Ito and Z. Kuznetsova and T. Tanaka and F. Toppan},
title = "{Integrable $\mathbb{Z}_2^2$-graded extensions of the {L}iouville and sinh-{G}ordon theories}",
journal = {J. Phys. A: Math. Theor.},
year = {2025},
month = {jan},
publisher = {IOP Publishing},
volume = {58},
pages = {055201},
doi = {10.1088/1751-8121/adaab3},
number = {5},
url = {https://doi.org/10.1088/1751-8121/adaab3},
}

@phdthesis{Ferreira,
  author = {J. M. C. Ferreira},
  title  = "{Generalized {B}{\"a}cklund transformations for {Toda} field theories}",
  school = {IFT-UNESP},
  year   = {2022},
  type   = {Ph.D. thesis}
}

@article{AFITTsuper,
  author = "{N. Aizawa and I. Fujii and R. Ito and T. Tanaka and F. Toppan}",
  title  = "{Integrable $\mathbb{Z}_2^2 $-graded super-{Liouville} equation and induced $\mathbb{Z}_2^2 $-graded super-{Virasoro} algebra}",
  journal = {},
  year = {2025},
  eprint = {arXiv:2512.17449[math-ph]}
}

@article{wang2025particle,
  title={Particle exchange statistics beyond fermions and bosons},
  author={Z. Wang and K. R. A. Hazzard},
  journal={Nature},
  volume={637},
  pages={314--318},
  year={2025},
  doi={10.1038/s41586-024-08262-7},
  publisher={Nature Publishing Group UK London},
  number={8045},
}

@article{toppanMulti,
  title={$\mathbb{Z}_2 \times \mathbb{Z}_2$-graded parastatistics in multiparticle quantum {Hamiltonians}},
  author={Toppan, Francesco},
  journal={J. Phys. A: Math. Theor.},
  volume={54},
  number={11},
  pages={115203},
  year={2021},
  doi={10.1088/1751-8121/abe2f2},
  publisher={IOP Publishing}
}

@article{toppan2021inequivalent,
  title={Inequivalent quantizations from gradings and $\mathbb{Z}_2 \times \mathbb{Z}_2$ parabosons},
  author={F. Toppan},
  journal={J. Phys. A: Math. Theor.},
  volume={54},
  pages={355202},
  year={2021},
  doi={10.1088/1751-8121/ac17a5},
  publisher={IOP Publishing},
  number={35},
}

@article{toppan2022first,
  title={First quantization of braided {Majorana} fermions},
  author={F. Toppan},
  journal={Nucl. Phys. B},
  volume={980},
  pages={115834},
  year={2022},
  doi={10.1016/j.nuclphysb.2022.115834},
  publisher={Elsevier}
}

@article{toppan2024detectability,
  title={On the detectability of paraparticles beyond bosons and fermions},
  author={F. Toppan},
  journal={Int. J. Geom. Methods Mod. Phys.},
  volume = {23},
  number = {6},
  year={2026},
  doi={10.1142/S0219887825400420},

}

@article{toppan2024braid,
  title={On braid statistics versus parastatistics},
  author={F. Toppan},
  journal={J. Phys.: Conf. Ser.},
  volume={2912},
  pages={012011},
  year={2024},
  doi={10.1088/1742-6596/2912/1/012011},
  number={1},
}

@article{huerta2025particle,
  title={Para-particle oscillator simulations on a trapped-ion quantum computer},
  author={C. Huerta Alderete and A. M. Green and N. H. Nguyen and Y. Zhu and N. M. Linke and B. M. Rodr{\'\i}guez-Lara},
  journal={J. Appl. Phys.},
  volume={138},
  pages = {054401},
  year={2025},
  publisher={AIP Publishing},
  doi ={10.1063/5.0276426},
  number={5},
}

@article{alderete2025experimental,
  title={Experimental realization of para-particle oscillators},
  author={C. Huerta Alderete  and A. M. Green and N. H.  Nguyen and Y. Zhu and B. M. Rodr{\'\i}guez-Lara and N. M. Linke},
  journal={Scientific Reports},
  volume={15},
  pages={41498},
  year={2025},
  doi = {10.1038/s41598-025-25271-2},
}

@article{polyakov1990gauge,
  title={Gauge transformations and diffeomorphisms},
  author={Polyakov, Alexander M},
  journal={Int. J. Mod. Phys. A},
  volume={5},
  number={05},
  pages={833--842},
  year={1990},
  doi={10.1142/S0217751X90000386},
  publisher={World Scientific}
}

\end{document}